\documentclass[pra,aps]{revtex4}
\usepackage{graphicx}
\input psfig.sty

\newcommand{\ba}{\begin{eqnarray}}
\newcommand{\ea}{\end{eqnarray}}
\newcommand{\be}{\begin{equation}}
\newcommand{\ee}{\end{equation}}
\newcommand{\bea}{\begin{eqnarray}}
\newcommand{\eea}{\end{eqnarray}}

\newcommand{\ket}[1]{|#1\rangle}

\begin{document}
\def\CC{{\rm\kern.24em \vrule width.04em height1.46ex depth-.07ex \kern-.30em
C}}

\title{Perfect initialization of a quantum computer of neutral atoms \\
in an optical lattice of large lattice constant}

\author{J. Vala$^{1,3}$, A. V. Thapliyal$^{2,3}$, S. Myrgren$^1$, U. Vazirani$^{2,3}$, D. S. Weiss$^4$ and K. B. Whaley$^{1,3}$}

\affiliation{
$^1$ Department of Chemistry and Pitzer Center for Theoretical Chemistry, University of California, Berkeley, California 94720\\
$^2$ Department of Computer Science, University of California, Berkeley, California 94720\\
$^3$ Mathematical Sciences Research Institute, 1000 Centennial Drive, Berkeley, CA 94720-5070 \\
$^4$ Department of Physics, Pennsylvania State University, University Park, Pennsylvania 16802-6300
}

\date{\today}

\begin{abstract}
We propose a scheme for the initialization of a quantum computer based on neutral atoms trapped in an optical lattice with large lattice constant. Our focus is the development of a compacting scheme to prepare a perfect optical lattice of simple orthorhombic structure with unit occupancy. Compacting is accomplished by sequential application of two types of operations: a flip operator that changes the internal state of the atoms, and a shift operator that moves them along the lattice principal axis. We propose physical mechanisms for realization of these operations and we study the effects of motional heating of the atoms. We carry out an analysis of the complexity of the compacting scheme and show that it scales linearly with the number of lattice sites per row of the lattice, thus showing good scaling behavior with the size of the quantum computer.
\end{abstract}

\maketitle

\section{Introduction}
\label{Sec:Int}

The present work is motivated by the possibility of using neutral (alkali) atoms trapped in an optical lattice for quantum computing.\cite{Deutsch:98,Jaksch:99,Jaksch:00}
We focus on a {\it scalable} quantum computing architecture where the number of qubits can in principle be increased using only polynomially more physical resources.
We consider one-dimensional (1-D), orthorhombic (in the near cubic arrangement) two-dimensional (2-D) and three-dimensional (3-D) lattices, although generalization of the proposed scheme to other structures is possible.
The lattice is characterized by a large lattice constant $a$ which allows each lattice qubit to be individually addressed during the computation.
The total number of sites $N$ is defined as the number of sites within the minimal compact volume enclosing the region of the lattice that is populated by atoms.
Assuming that the lattice sites are either occupied by a single atom or are vacant, we define a filling factor $f = N_{occ}/N$, where $N_{occ}$ is the number of (singly) occupied sites. 
%We shall refer to the occupation as the map of atom locations.
Under the assumption that all lattice sites are either occupied with a single atom or empty
(i.e. with a uniform distribution over the lattice), the filling factor
is the same as the site occupancy probability $p_{occ}$.

We investigate here the preparation of the system for quantum computation, i.e. its {\it initialization} as a quantum computer.
Our key objective is a perfect optical lattice where each lattice site is occupied with a single atom in a specific internal state and the motional ground state.
We present a feasible and scalable scheme for {\it compacting} the optical lattice to achieve this state with 100\% fidelity. 
Our scheme removes the vacant sites to the edge of the initial lattice $(N_i,f_i)$ and thus creates a smaller but perfect final lattice $(N_f,f_f)$, where $N_f < N_i$ and $f_f = 1 > f_i$.
This compacting procedure can be used for
initialization, or for reinitialization of a quantum computer before a new computation.
Other applications may also be found in the construction of fault-tolerant procedures for the computation itself.

A motivation for developing the compacting procedure is the conjecture that randomly occuring imperfections, even if their locations are known, cause bottlenecks in quantum information flow which can not be avoided when the computer size, i.e. the number of lattice qubits, scales up.
In fact, we maintain that the probability of finding a ``good'' sublattice is exponentially small for any constant filling factor.
This is because the probability that there will be 'insurmountable' blocks of defects (gaps) in any chosen sublattice increases rapidly with its size. The site percolation threshold $p_c$ reaches the following values: 1 (1-D), 0.59 (2-D) and 0.31 (3-D) \cite{Hughes:95,Grimmet:99}. Thus the initial filling factor $f_i = 0.5$ exceeds the percolation threshold only in a three dimensional lattice.
Even in this case, the condition $f_i > p_c$ only implies a non-zero probability that distant qubits may be connected. It does not ensure that they are connected via independent routes, nor even that they are connected at all.
In addition, the mapping of a quantum algorithm onto an imperfect lattice may be a hard classical computational problem (even NP hard).
Based on these considerations, an imperfect lattice structure seems to pose a stumbling block in the realization of a scalable quantum computer.

We briefly list other possible approaches to preparing an optical lattice with single occupancy at each site.
The quantum phase transition from the superfluid to Mott insulator phase \cite{Fisher:89,Oosten:01}, recently observed in \cite{Greiner:02}, can prepare a singly occupied optical lattice when both the lattice constant and the well depth are rather small ($a \approx$ 0.5 $\mu$m, $U_0 \approx$ 1 $\mu$K).
The quantum critical point is given by the ratio of on-site atomic interaction energy $U$ and the tunneling strength $T$ as $U/T = 10.6 ~d$, where $d$ is the dimensionality of the lattice \cite{Oosten:01,Greiner:02}.
With increasing values of lattice constant or well depth, the tunneling strength diminishes exponentially, bringing the system deeply into the Mott insulator phase.
Here, the band structure of the lattice potential energy spectrum approaches its discrete limit, and is suitable for robust quantum computation with independent qubits.
This approach can prepare a singly occupied optical lattice of small lattice constant with more than 90\% of fidelity. However, it is difficult to apply it to an addressable optical lattice, where the large lattice constant sets an unrealistically long timescale for adiabatically changing the lattice.
In addition, even if the phase transition could be established, the final singly occupied optical lattice would be unlikely to reach the desired level of perfection, and subsequent compacting would be required.
Another recent proposal for preparing a nearly perfect optical lattice is based on use of the dipole interaction between atoms excited to Rydberg states \cite{Jaksch:00,Lukin:01}.
Here, quantum state manipulations exploiting the dipole blockade effect can lead to a lattice with a significantly fewer vacant sites \cite{Saffman:02}.
Neither of the two previously presented approaches can currently guarantee a perfect lattice where each site is occupied by exactly one atom. 
An alternative proposal for preparing a perfect optical lattice has recently been proposed.
It involves adiabatic loading of one optical lattice from another that has one or more atoms at every site \cite{Rabl:03}.

\section{Physical system}
\label{Sec:Phy}

%\subsection{Optical lattice}
%\label{Sec:Opt}

An optical lattice with individually addressable sites is a particularly promising candidate for implementation of quantum computation. 
This important requirement is satisfied by having a large lattice constant ($a \approx 5~\mu m$).
A 1-D optical lattice can easily be realized by interfering two linearly polarized laser beams \cite{Scheunemann:00}. Higher dimensional lattices can be implemented using two (2-D) or three (3-D) pairs of laser beams in a mutually perpendicular arrangement. 
Undesired interference between sublattices can be eliminated by using slightly different frequencies for each of the beam pairs. 
The resulting lattice possesses a simple orthorhombic structure.
Such an optical lattice with $a \approx 5~\mu$m can be made with CO$_2$-laser beams \cite{Scheunemann:00} or with blue-detuned light. The blue detuned standing waves consist of two beams propagating at a shallow angle $\theta_b$ with respect to each other, giving a lattice spacing $a = \lambda / 2sin(\theta_b/2)$.

A pair of counterpropagating (along the $z$ axis) linearly polarized laser beams of identical wavelength $\lambda$ gives rise to a superposition of standing waves of opposite circular polarization $\bf{e_{\pm}}$. The electric field is given by:

\be \label{Eq:Field}
E(z) = \sqrt{2}E_0[e^{i\theta /2} cos(kz - \theta /2){\bf{e_{-}}} - e^{-i\theta /2} cos(kz + \theta /2){\bf{e_{+}}}].
\ee
Here $\theta$ is the relative angle between the linear polarization vectors of both beams, $k = 2 \pi / \lambda $, and $E_0$ is the single beam field amplitude.
In the absence of any additional external field, the resulting 1-D periodic lattice potential depends on the magnetic hyperfine sublevel \cite{Deutsch:98}. It can be characterized by the following relation
\be \label{Eq:Latt}
U(z) = \frac{U_0}{2} cos(\theta ) cos( 2 k z) + \frac{U_1}{2} sin(\theta) sin( 2 k z),
\ee
where $U_0 = \frac{2}{3} \tilde{\alpha} E_0^2$ and $U_1 = \frac{1}{3} \tilde{\alpha} E_0^2 \frac{m_F}{F}$ describe the well depths of the potential at $\theta = 0$ and $\theta = \pi$ respectively, $\tilde{\alpha}$ is the characteristic polarizability of a given transition, $F$ is the total angular momentum of the relevant atomic hyperfine level and $m_F$ is the magnetic hyperfine sublevel. 
Note that the potentials for all magnetic hyperfine sublevels coincide for $\theta = 0$. 

One-dimensional as well as multi-dimensional orthorhombic arrangements with approximately 20 sites per dimension and 150 $\mu$K depth are readily achievable.
For 1-D, 2-D and 3-D lattice potentials and a filling factor of 0.5, this results in 10, 200, and 4000 atoms, respectively. 
We focus on an optical lattice filled with atoms of $^{133}$Cs, although
the proposed scheme is applicable to other alkali atom.
The $6s ^2S_{1/2}$ electronic ground state of $^{133}$Cs consists of two hyperfine levels of total angular momentum $F = 3$ and $F = 4$, with energy splitting $\Delta E = E_{F=4} - E_{F=3} = 9.1926$ GHz.

\section{Loading and cooling of optical lattice}

A magneto-optical trap \cite{Raab:87} can be used to load well-spaced far-off-resonant optical lattice sites with many atoms (N=10-100) \cite{Winoto:99}.  Atoms are lost during laser cooling in the lattice, in pairs, due to photon assisted collisions \cite{DePue:99,Schlosser:01}.  Eventually, the sites initially occupied by an odd number of atoms become singly occupied, and initially even occupied sites are left empty.  The filling factor, $f_i$, that results from this combined loading and cooling process approaches a maximum of 0.5.  This sort of loading of a far-off-resonant optical lattice with Cs atoms, to 44\% filling, has been demonstrated experimentally with a small lattice constant ($\sim 0.5~\mu$m).

Trapped atoms can then be brought to the vibrational ground state of their lattice sites using Raman sideband coolling \cite{Kerman:00,Hamann:98,Vuletic:98,Han:00,Perrin:98}.  This procedure has been experimentally demonstrated in 1D \cite{Vuletic:98,Perrin:98}, 2D \cite{Hamann:98} and 3D \cite{Kerman:00,Han:00} optical lattices.  In \cite{Han:00}, the 3D ground state was populated by up to 55\% of atoms, corresponding to over 80\% in each dimension.  It was shown that the cooling was limited by the rescattering of cooling photons.  This mechanism is dramatically reduced at the low densities associated with large lattice constants.  Even with a 5 $\mu$m lattice constant, a ~150 $\mu$K depth lattice leaves the atoms well within the Lamb-Dicke limit required for efficient sideband cooling.  It is reasonable to expect nearly 100\% vibrational ground state population after Raman sideband cooling.

With a 150 $\mu$K deep, 5 $\mu$m spaced lattice, the vibrational level spacing is  well below the expected minimum polarization gradient cooling temperature of ~2 $\mu$K, so the polarization gradient cooling limit should be obtained \cite{Winoto:99}.  With a temperature so small compared to the lattice depth, there will be negligible site hopping during cooling.  Therefore the atoms can be imaged during the cooling for as long as needed. 1D CO$_2$ lattices have recently been imaged \cite{Scheunemann:00}.  In a similar way, it should be possible to use $\sim$0.6 numerical aperture optics to image successive 2D planes, and thus determine the site occupancy of a 3D lattice.  Since the localized atoms in the 3D lattice will scatter light coherently, there will be significant interference between light scattered from atoms in the image plane and light from non-imaged planes.  Information can be obtained from such signals, easier so if the lattice spacing can be made an integer multiple of image wavelengths, as is possible in a blue-detuned lattice.  The total intensity of light from each non-imaged plane is only  about 2\% of the intensity from the image plane, making the signal to background from a 4000 atom 3D lattice $\sim$3.  When the lattice occupation is perfectly known, it can then be compacted into a perfectly occupied lattice.

\section{Compacting procedure}

\begin{figure}
\includegraphics[width=3.0in] {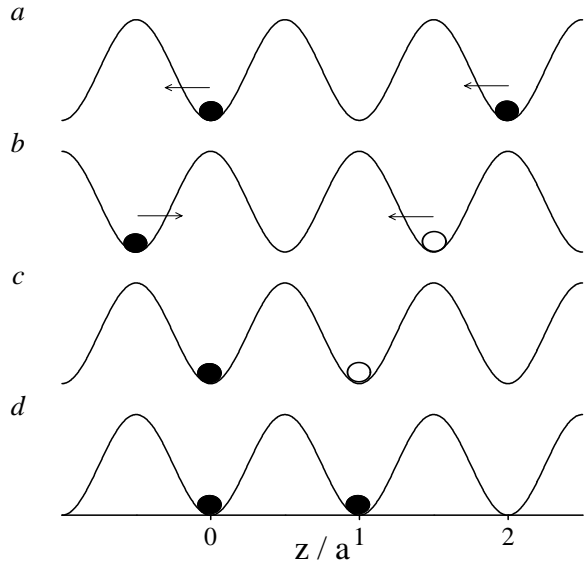}
\caption{Compression scheme. a) All atoms of the lattice are in a single internal state and are shifted by $+a/2$ as a result of rotation of the relative polarization angle $\theta$  from $0$ to $\pi$. b) The state of the mobile atoms is then flipped, e.g. from $m_F = +1$ (filled circles) to $m_F = -1$. This is accompanied by a change in shift direction, denoted by the arrows. c) Rotation of the angle $\theta$ back from $\pi$ to $0$ moves  the $m_F = +1$ atom back to its original position and the $m_F = -1$ atom forward to the next lattice site, providing the desired compacting of the lattice structure. d) The mobile atoms are then flipped back to the original state. The net effect is a shift of the right hand atom by one site to the left, or equivalently, of the corresponding vacancy by one site to the right.}
\label{Fig:Scen}
\end{figure}

We now present a scheme for converting an imperfect optical lattice, where each site is either empty or occupied by a single atom, into a smaller lattice where each site is occupied by a single atom.
Our scheme exploits the ability to move a subset of trapped atoms to fill vacant sites, thereby removing vacancies to the edge of the lattice potential.
Compacting a lattice is equivalent to sequentially removing vacant sites to lattice edges.

We define two elementary operations that are sufficient to compact the lattice. They are (1) the flip operator, $\hat{F}_{I}$, which toggles the state of an atom at position $I = (i,j,k)$ between two levels, and (2) the shift operator, $\hat{S}_{IJ}$, which moves an atom from position $I$ to a neighboring position $J = (i',j',k')$, where only one of the coordinates of $J$ differs from $I$ by one. 

Compacting results from translating some of the atoms, which we call {\it mobile} atoms,  to fill vacant sites, while the rest, which we call {\it stationary} atoms, remain fixed within their lattice structure.
We refer to the state of a mobile atom as the {\it mobile} state, and that of a stationary atom as the {\it storage} state.
These may for instance be $|Cs ~6S_{1/2},F=3,m_F=-1\rangle$ and $|Cs ~6S_{1/2},F=3,m_F=+1\rangle$ respectively.
The mobile atoms are initially selected based on the map of lattice occupancy obtained from the imaging process.
The compacting process starts with all atoms in the storage state.
Atoms are moved by rotating the linear polarization of one of the lattice beams by angle $\theta$ relative to the counterpropagating beam, according to the following steps:
\begin{enumerate}
\item{for $\theta = 0 \to \pi$, all atoms are initially in the storage state and all move by $-a/2$, where $a$ is the lattice constant and  the sign indicates direction; formally, this operation keeps the lattice occupation structure invariant and hence acts on it as identity;}
\item{at $\theta = \pi$, the mobile atoms are selectively flipped into the mobile state, while the states of stationary atoms remain unchanged;}
\item{for $\theta = \pi \to 0$, the stationary atoms are shifted back to their original positions ($+a/2$) while the mobile atoms are shifted forward to the next vacant lattice site ($-a/2$);}
\item{the mobile atoms are selectively flipped back to the storage state.}
\end{enumerate}
After step 3, the lattice occupation structure is now different. All atoms that were selected to be mobile have been shifted to vacant sites. Vacancies have been displaced to the edge of the lattice. The procedure is repeated until all vacant sites have been moved to edge of the lattice. This scheme is summarized in Fig. \ref{Fig:Scen}.
For practical implementation, the scheme can be simplified by not requiring that storage atoms end each operation in the same position. 
As the lattice is compacted to its center, translations of the storage atoms will tend cancel each other out.

In the next two sections we consider the physics of each elementary step in detail in order to place bounds on the extent of undesired heating.

\section{Flip operation}
\label{Sec:Ato}

\subsection{Achieving site selectivity}

The compacting scheme requires that we can make an atom at a single site undergo a flip transition, while none of its neighbors make the transition. We propose two distinct ways to accomplish this feat.  The first, and more novel, approach uses an independent ``addressing'' beam, which is a far-off-resonant, circularly polarized laser beam tightly focused on the atom to be flipped.  The circularly polarized beam shifts the stationary and mobile atomic states differently, so that the resonance transition between them is shifted.  The addressing beam will have non-zero intensity at non-target atoms, especially those than lie along the addressing beam axis.  But as long as the Rayleigh range is reasonably much smaller than the lattice spacing, the resonance frequency shift will be much smaller at non-target sites.  In the presence of the addressing beam, a pulse from a spatially homogeneous source can be used to flip the target atom.  The requirement is that the pulse time be long compared to the inverse of the difference between the resonance frequency shifts of the target and the nearest non-target atoms.  In that case, non-target atoms will be far enough off resonance that they will not be flipped.  These flip pulses can be driven either by direct microwave excitation or by co-propagating stimulated Raman beams. The required parameters of an addressing beam are easily attained.  For instance, a 2 $\mu$W laser beam at 877 nm, focused to a waist of 1.2 $\mu$m, gives a 1 MHz relative Stark shift, 2.5 times larger than that of the nearest neighbor.

Another way to get site selectivity is to drive an off-resonant stimulated Raman transition using tightly focused perpendicular laser beams, so that only atoms at the target site see appreciable intensity from both beams.  An advantage to this approach is that, since there is no need to frequency resolve the transitions at different sites, the flip can be accomplished much more quickly.  We will discuss some disadvantages in the following subsections.  

\subsection{Driving flip transitions}

All of the various ways to make site-selective flips have the option of driving a $\pi$-pulse or using adiabatic fast passage.  With $\pi$-pulses, the best approach is probably to tailor the pulse shape, as with a Blackman pulse, in order to minimize off-resonant excitation.  The option also exists to use a square pulse, but it requires to make sure that all non-target atoms lie close to the first minimum of the resulting sinc function \cite{Kasevich:92}.  Either way, $\pi$-pulses require very stable and repeatable field intensity.  In the case where Raman beams are used without an addressing beam, stable and repeatable centering of the tightly focused beams on the target atom would be a significant challenge for making a reliable flip.

\begin{figure}
\includegraphics[width=3.0in] {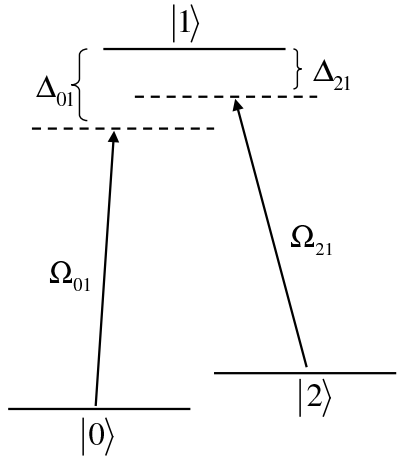}
\caption{The three-level quantum system coupled by two optical fields in a $\Lambda$-configuration.
$\Omega_{ij}$ is the Rabi frequency for coupling of the states $i$ and $j$, $\Delta_{ij}$ is 
the detuning from a resonance $\omega_{ij} = E_i - E_j/\hbar$.}
\label{Fig:Lambda}
\end{figure}

The flip operation can also be implemented using rapid adiabatic passage. 
Consider a three level optical system in a $\Lambda$-configuration (see Fig. \ref{Fig:Lambda}) where two ground states $|0\rangle$ and $|2\rangle$ are coupled indirectly via an intermediate state $|1\rangle$, using a pair of laser fields. 
Adiabatic passage occurs when the initial and target levels coupled off-resonantly through the intermediate state sweep through their dressed-state resonance as result of interaction with a linearly frequency chirped field (for a description of the chirped field and its interaction with matter see for instance \cite{Vala:01a,Vala:01b}).

The atom-field Hamiltonian is (in the interaction representation)
\be \label{Eq:GHam}
{\bf {\hat H}}(t) = {\bf {\hat H}_0} + {\bf {\hat H}_I}(t) = \sum_{i=0}^{2} \hbar \omega_i \vert i \rangle \langle i \vert + [ W_{01}(t) \sigma_{01}^+ + h.c. ] + [ W_{12}(t) \sigma_{12}^+ + h.c. ]
\ee
where $\sigma_{ij}^+ = \vert j \rangle \langle i \vert$ and $W_{ij}(t)$ is the time-dependent coupling strength given by the external optical field.
Canonical transformation of this system with coupling fields detuned from resonance with the intermediate state results in an effective two-level system \cite{Madelung:78,Imamoglu:99}:
\ba
{\bf {\hat H}_{eff}}(t) & = & {\bf {\hat H}_0} + {\bf {\hat H}_{Stark}} + {\bf {\hat H}_{I,eff}}(t) \nonumber \\
& = & \sum_{i=0}^{2} \hbar \omega_i \vert i \rangle \langle i \vert \nonumber \\
& + & \frac{\vert W_{01}(t) \vert^2}{\Delta_{01}} \sigma_{01}^z + \frac{\vert W_{12}(t) \vert^2}{\Delta_{12}} \sigma_{12}^z \nonumber \\
& + & W_{02} \sigma_{02}^+ + W_{02}^{*} \sigma_{02}^-
\ea
where $W_{02} = \frac{W_{01}W_{12}}{2}(\frac{1}{\Delta_{12}}-\frac{1}{\Delta_{01}})$ is the effective coupling strength between the states $\vert 0 \rangle$ and $\vert 2 \rangle$, and $\Delta_{ij}$ is the detuning of the field from the atomic transition between levels $i$ and $j$.
Application of a linearly chirped field for one of the transitions $0 \to 1$ or $2 \to 1$ results in a robust and complete flip operation \cite{Vala:01b}.
Such an adiabatic rapid passage can also be done with microwave radiation.

\subsection{Heating due to flip transitions}
      
      Site selective flip operations can cause heating in several distinct ways. The first is due to the impulse that both target and non-target atoms can feel during the process of turning on the spatially inhomogeneous beams, including either the addressing beam or the tightly focused stimulated Raman beams.  However, when the frequency of these beams is tuned between the first excited state fine structure levels ($6P_{1/2}$ and $6P_{3/2}$ in Cs) the ac Stark shifts due to the two levels are opposite.  For any individual ground state sublevel, there is a magic frequency where the two Stark shifts cancel. To avoid this heating, one simply needs to use the magic frequency for the storage hyperfine sublevel, for instance, 877 nm for the $F=3$, $m_F=1$ hyperfine sublevel.  The difference in frequency between the Raman beams is negligible on the scale needed to avoid this heating effect.
	
A second heating mechanism only applies when an addressing beam is used, and it results because the addressing beam necessarily Stark shifts the mobile sublevel.  Therefore the trapping potential of an atom in that state is the sum of that due to the optical lattice light and the addressing light.  The vibrational frequencies are thus different for the two hyperfine sublevels. Flip transitions between the storage and mobile sublevels can be made in two limits.  If the pulse time is long compared to the inverse of the vibrational state splitting, so that the vibrational states are resolved, then the atom can make a transition to the new vibrational ground state, and there will be no heating.  If the pulse time is shorter than the inverse of the vibrational state splitting, then the original vibrational wavefunction will be projected onto a superposition of states in the new basis.  These levels will tend to dephase, and the atom will no longer be in the vibrational ground state when it is returned to the storage state.  This heating is significantly reduced when the addressing beam is weakened, which requires that the microwave pulse be longer.  Calculations of this heating effect have been performed in the context of making single qubit operations in a site addressable optical lattice, and the heating is $\sim 10^{-4}$ vibrational energies per 30 $\mu$s flip operation \cite{Vaishav:XX}.
      
Another heating mechanism is due to the photon recoil from the flipping radiation.  In the case of microwaves and co-propagating Raman beams, the photon recoil is negligible. But for the orthogonal Raman beams, unless the pulse is slow enough to resolve the atom's vibrational states, the target atom will receive a significant photon recoil kick.  In the Lamb-Dicke limit, the probability of vibrational excitation per pulse is not high, but this would still likely be the dominant source of heating in the compacting sequence.

\section{Shift operation}
\label{sec:shift}

The shift operator moves the lattice trapped atoms over 
a distance $a/2$, where $a$ is the lattice constant.
As mentioned earlier, all the atoms are initially prepared in a storage
state, e.g. $m_F=+1$. 
It is easily seen from Eq. (\ref{Eq:Latt}) that the atoms will move
when the linear polarization vector of one of
the lattice beams is rotated.
Rotating the relative angle of polarization $\theta$ 
by $\pi$ moves all the atoms by $a/2$.
Switching the internal state of the mobile atoms to the mobile state, and
slowly returning the polarization to its original value, $\theta = 0$,
returns the stationary atoms back to their original positions, while each
mobile atom is moved forward another $a/2$, so that it is separated
by $a$ from its original location. Finally, the mobile atoms are returned to 
the storage state.

As seen from Eq. (\ref{Eq:Latt}), at $\theta=0$ the atoms are confined only by
the first term of the potential. As $\theta$ increases, the second
term starts to dominate the optical lattice potential. 
At $\theta=\pi /2$, where the well-depth is $U_1$, only this state sensitive term confines
the atoms. Further rotation to $\theta=\pi$ increases the well depth back to its original value, $U_0$.
We assume that the relative angle between the polarization vectors
of the lattice beams, $\theta$, is a linear function of time, i.e. $\theta = \beta t$,
though it is conceivable that a non-linear variation may be useful to further reduce
the heating under certain conditions.

\begin{figure}
\includegraphics[width=3.0in] {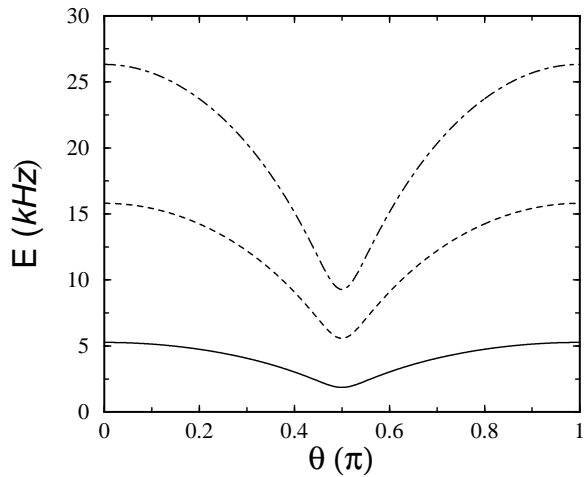}
\caption{Change of the lowest vibrational eigenstates of the periodic potential for the atomic state $|Cs ~6S_{1/2},F=4,m_F=-1\rangle$ as a function of the rotation of the relative polarization angle $\theta$. The energy is measured relative to the potential minimum. The vibrational frequency for the transition $0 \to 1$ is $10.5 ~kHz$ for $\theta = 0$, and $3.66 ~kHz$ for $\theta = \frac{\pi}{2}$. }
\label{Fig:Omega}
\end{figure}

The highest vibrational heating occurs for the largest variation of
the potential during the shift operation.
For ground state Cs atoms with $|m_F| = 1$ in a 50 nm blue detuned optical lattice, 
the potential well-depth changes by a factor of approximately $U_0/U_1 = 8$ during one shift operation.
This ratio is higher in the case of the $CO_2$ optical lattice, which is much further
detuned from the atomic transition.
The corresponding change in the frequency of the lowest vibrational levels
is illustrated in Fig. \ref{Fig:Omega}.
We now analyse the case of 
$U_0 = 100 ~\mu K$ and $a = 5.3~\mu m$, in order to obtain an
upper bound on the heating process.

\subsection{Vibrational heating}
\label{subsec:vibheat}

Initially each lattice atom is in its vibrational ground state with energy $U_0$.
The shift operation displaces the lattice potential and causes 
vibrational excitation of atoms. This heating can be quantified by
the total energy of a moving particle relative to the initial state, $<E> - ~U_0$.
We have employed both analytical and numerical techniques to get insight into
the dynamics of atoms in a time-varying lattice potential.
Application of the analytical approach is limited to the {\it simplified case} where both
terms of the potential (\ref{Eq:Latt}) are of the same magnitude.
Numerical simulation was made using a Fourier
grid representation of quantum states and operators and
the Chebychev polynomial quantum propagation method. Details of the simulation methods can be
found in Appendix \ref{App:Methods}. The explicit time-dependence of
the potential was approximated in discrete steps, with a 
time step $\Delta t \approx 1/U_0$ chosen to 
ensure the accuracy of the final results to within $\sim 10^{-5}$.

\subsubsection{$U_0 = U_1$ case}

\begin{figure}
\includegraphics[width=3.0in] {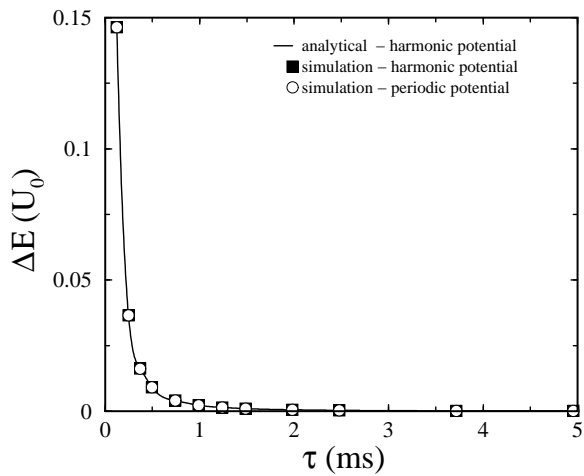}
\caption{The maximal vibrational heating of the initial ground state population $\Delta E$ as a function of the total timescale $\tau$ of the shift operation. The analytical and numerical simulation results in the harmonic approximation are compared together and with the numerical simulation using the simplified periodic potential (see the text for explanation). The analytical and numerical results are indistinguishable on this scale.
\vspace{1cm}}
\label{Fig:ShiftE}
\end{figure}
\begin{figure}
\includegraphics[width=3.0in] {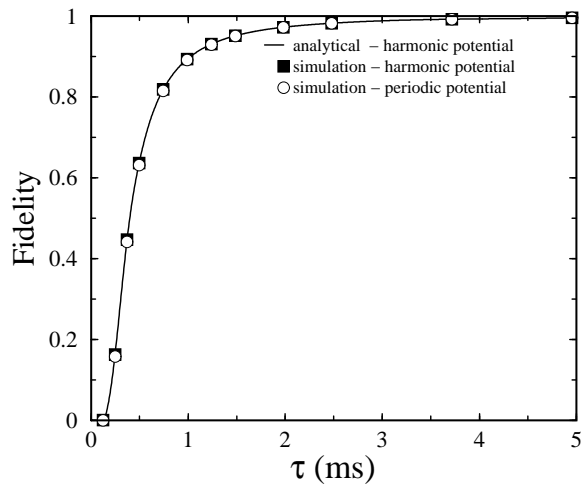}
\caption{The minimal fidelity $\tilde{F}$ as a function of the shift operation timescale for the simplified case (see the text for explanation).
The very good agreement between the analytical and numerical results proves that the classical picture of the process based on the coherent state representation correctly reproduces the atomic dynamics induced by the time-varying potential. For short timescales, the anharmonicity of the periodic potential slightly reduces the fidelity compared to the results obtained within the harmonic approximation.}
\label{Fig:ShiftF}
\end{figure}

We first consider the simplified case where the two terms in Eq. (\ref{Eq:Latt}) have the same magnitude, i.e. $U_0 = U_1$.
In the vicinity of the minimum, we can approximate the periodic potential of 
the lattice with a second order Taylor expansion, yielding the harmonic vibrational frequency
$\omega = k \sqrt{2U_0/m}$, where $k = 2\pi/\lambda$.
Comparison with the results of direct diagonalization
of the periodic potential shows that the fractional difference from the true frequency is
$\sim 10^{-3}$, and that the anharmonicity
of the periodic potential (manifested in the deviation of $\omega_{k,k+1} = (E_{v=k+1} - E_{v=k})/\hbar$
from the harmonic frequency) is linear in the vibrational quantum number 
$k$ and is only ~3 \% for $k = 20$.
These facts justify the use of the harmonic approximation to get analytical insights
into the process of vibrational heating.

The motion of the simplified potential, induced by rotation of the polarization vector
of one of the lattice beams, is linear in time and induces a transfer of the energy
into the system
$E = m v^2 = m \frac{\Delta z^2}{\tau^2}$.
Here $m$ is the atom mass, and $\Delta z$ is the total distance of the potential translation 
over the duration $\tau$.
E is the most energy that can be transferred into vibrational motion during the beginning or end of a translation.
Onset of the potential translation at time $t = 0$ causes displacement of the initially stationary
vibrational state
with respect to the moving potential reference frame.
The displacement transforms the stationary initial state into a vibrating coherent state with
a maximal displacement of $x_{max}$. 
This displacement is related to the transferred energy
via $E = \frac{1}{2} m \omega^2 x_{max}^2$, so
$x_{max} = \frac{\sqrt{2}}{\omega}\frac{\Delta z}{\tau}$.
The maximal energy in the system at $\theta = \pi$
is $2 E$. The analytical and numerical results for the harmonic and periodic
potential show perfect agreement, as illustrated in Fig. \ref{Fig:ShiftE}, and hence
justify the applied harmonic approximation and the coherent state representation.

To further characterize the quality of the shift
operation, we define a fidelity measure,
$\tilde{F}=|\langle \Psi_i | \Psi_f \rangle|^2 = |\langle v=0 | \Psi_f \rangle|^2$,
which corresponds to the projection
of the final coherent state onto the vibrational ground state of the translated potential.
In this simplified case, the fidelity can be evaluated analytically, yielding 
$\tilde{F} = exp(- \frac{m \omega}{4 \hbar} x_{max}^2) = exp(-\frac{m}{2 \omega \hbar}\frac{\Delta z^2}{\tau^2})$.
As can be seen in Fig.~\ref{Fig:ShiftF}, the fidelity remains low for shift operation
times less than $0.5 \; ms$, whereafter it rises sharply. For shift times
longer than $2 \; ms$ the fidelity is very close to unity, and
thus provides a good estimate of the timescale necessary to preserve
adiabaticity.

\subsubsection{Control of heating}

\begin{figure}
\includegraphics[width=3.0in] {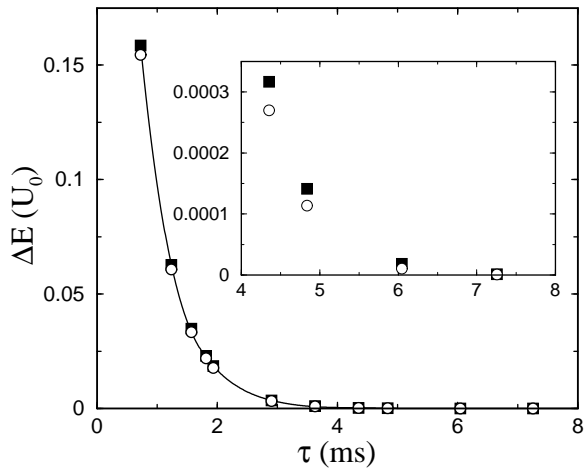}
\caption{The maximal vibrational heating of the initial ground state population $\Delta E$, as a function of the total timescale $\tau$ of the shift operation on the realistic potential (see the text for explanation). The heating has two contributions. The larger component originates from acceleration and deceleration of the potential during the motion. The smaller component comes from the phase of the vibrational motion of the coherent state at the point when the potential stops ($\theta = \pi$) and it is controllable by suitable timing of the shift operation as in the simplified analysis (see text).\vspace{1cm}}
\label{Fig:ShiftER}
\end{figure}

Exploration of this simplified model provides insight into the
control and minimization of vibrational heating.
The onset of the potential motion generates a coherent state from the initial
vibrational ground state. In the moving reference frame, the potential motion first displaces
the initial state to the negative momentum and negative coordinate region of the phase space,
while setting its motion along the classical phase space trajectory corresponding to
the coherent state.
At the final instant of time when $\theta = \pi$, the actual motional phase of the coherent state
accumulated during the evolution ($e^{-i\omega \tau}$) 
determines whether stopping the potential will reduce or increase the total vibrational
energy of a trapped atom.
If the shift operation duration $\tau$ is an integer multiple of the vibrational period, then
the deceleration of the potential cancels the heating from the acceleration.
It is thus possible to control and minimize the heating from this process.

\subsubsection{Realistic case}
\label{subsec:adcond}

We next consider the {\it realistic case} where the depth of the potential depends
on the relative linear polarizations of the lattice beams. For our parameters, the depth changes
from $U_0$ to $U_0/8$ and back to $U_0$, when the angle between polarizations is rotated
through $\pi$ radians. The vibrational frequency is therefore also modulated,
accelerating the atoms when $\theta < \pi/2$ and deccelerating them when $\theta > \pi/2$.
This process of acceleration and deceleration of the motion, absent in the simplified 
analysis above, dramatically affects
the final amount of vibrational heating of the atoms in the realistic potential.

The real system can also heat due to the motional phase effect discussed
in the simplified case. As in the simplified case, it can be eliminated here
by suitable timing of the shift operation.
Although for the case $U_0/U_1 = 8$, it is a rather small portion of the total heating
(see Fig. \ref{Fig:ShiftER}), it becomes significant when either the duration of the shift
increases or the ratio  $U_0/U_1$ decreases (which ratio depends on lattice detuning
and the choice of atomic states.

\begin{figure}
\includegraphics[width=3.0in] {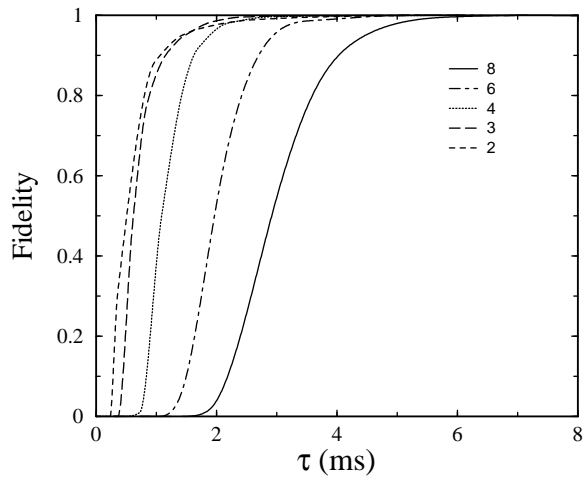}
\caption{Plots of the minimal fidelity $\tilde{F}$ as a function of the shift operation timescale for the full potential with time varying ratio $U_0/U_1$. As this ratio decreases, the contribution to the vibrational heating from the acceleration of the potential motion is decreasing, and eventually becomes dominated by the oscillatory contribution. The latter has the same origin as that in the simplified analysis (no acceleration) and is fully controllable by suitable timing of the shift operation. The character of the fidelity plot in this case also shows slower convergence to $\tilde{F} = 1$ typical for the simplified case (Fig. \ref{Fig:ShiftF}). This indicates that for durations $\tau \approx 3~ms$, a higher fidelity can be obtained with the potential $U_0/U_1 = 3$ than for $U_0/U_1 = 2$.\vspace{1cm}}
\label{Fig:ShiftFR6}
\end{figure}
\begin{figure}
\includegraphics[width=3.0in] {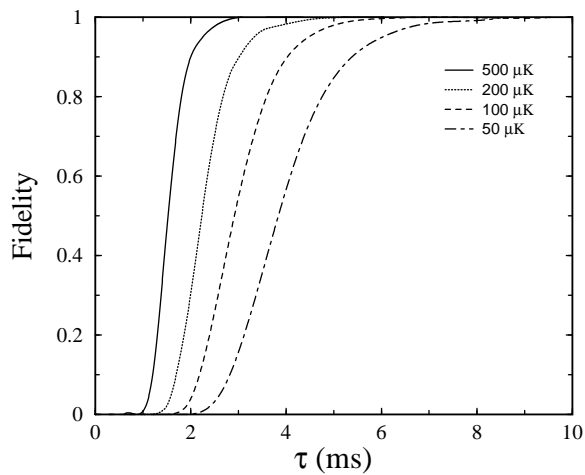}
\caption{The minimal fidelity $\tilde{F}$ as a function of the well depth $U_0$  for the realistic case with $U_0/U_1 = 8$.\vspace{1cm}}
\label{Fig:ShiftFU}
\end{figure}

The presence of two contributions to the heating energy is apparent from the fidelity plots
shown in Fig \ref{Fig:ShiftFR6}. If the acceleration dominates the heating, the fidelity converges faster
to its maximal value $\tilde{F} = 1$ compared to the case dominated by the controllable (oscillatory)
contribution.

Optimal conditions can be achieved by varying 
time duration $\tau$, well-depth $U_0$ (see Fig. \ref{Fig:ShiftFU}) and the ratio $U_0/U_1$.
When $U_0/U_1 = 8$, 
the atoms can be shifted on a time scale of $5 - 6 ~ms$
without appreciable heating, as shown in Fig.~\ref{Fig:ShiftER}. Moreover it can be seen that
even for fairly fast shift operations, $\tau < 2 ~ms$, the
atoms remain deeply trapped in the lattice, although recooling of the lattice may be required
after many operations.
The fidelity $\tilde{F}$ provides a sensitive measure of error.
Figure \ref{Fig:ShiftFR6} shows that atoms usually stay
in the ground state when $\tau > 6 ~ms$.

\subsubsection{Scaling}

Given that the initial lattice filling factor is only $f_i = 0.5$,
multiple shift operations will certainly be required to compact the lattice.
Thus, it is important to ensure that the heating of the atoms is
a {\em well-behaved} function of the number of shifts performed.
In fact, it can be expected that the maximal heating
scales as a linear function of the number of the shift operations, because of
additivity of energy. There is no reason to expect a different scaling with
the number of shifts if these are uncorrelated. 
When an enlargement of the computer size requires an increase of the number of the compacting
operations beyond the tolerable heating limit,  an intermediate recooling of
the lattice by Raman sideband cooling can be applied to reset the ground
state occupation to one.

\section{Efficiency of Compacting}
\label{Sec:Complexity}

To systematically analyze the efficiency of compacting,
we in turn consider finite 1-D, 2-D, or 3-D 
lattices with 50\% initial site occupation $p_{occ}$.
From imaging during laser cooling, we have a map of the lattice occupation.
Our goal is to move the atoms so that they form a single
contiguous block with no vacancies, i.e., a perfect finite lattice.
The primitive operation in this analysis is the shift of an atom
or group of atoms by one lattice site. 
We define the cost (time) for this operation to be one unit.  The physical
implementation of this operation has been discussed in
section \ref{sec:shift}, where we showed that moving a group of $n$ atoms through
one lattice site requires two movements of the trapping potential (2 elementary shifts) and
$2n$ flip operations (see Fig. \ref{Fig:Scen}). 
Note that there are only two elementary shifts because all atoms
move in a shift.
For fault tolerant computation \cite{Nielsen:00} we need to have
$O(n)$ parallel operations \cite{Note:1}, so for evaluating the scalability of the scheme
we assume that the spin flips can be achieved in parallel.
This is a good assumption, since the time required for a spin flip is much
smaller (by a factor of 100-1000) than the time to move the
trapping potential. In fact, we can ignore the spin flip cost altogether.
We also ignore the cost of classical computations required
to plan and implement the compacting procedure since, with proper
optimization, this cost scales at most linearly with the total
number of lattice sites. Therefore, the total cost is essentially
determined by the shift operations. 

We first study the 1-D lattice to show the solution in a simple
setting.  Then we consider the 2-D lattice, which presents
additional challenges.  Finally, we generalize the techniques of
the 2-D lattice to the 3-D lattice.

\subsection{One Dimensional Lattice} The one dimensional lattice
$L_n$ has $n$ sites.  The site occupation probability is $p_{occ}$.  Our
goal is to move the atoms so that they form a line of atoms with
no gaps in between.

Here, we use the obvious algorithm, which we call {\bf COMPACT},
to remove the vacancies.  The algorithm moves the atoms to the
left so that all the vacancies move to the right.  Thus after
running the algorithm we end up with a line of atoms at the left
of the lattice.

\noindent{\bf COMPACT:}
\begin{enumerate}
\item  Find the leftmost vacancy $v$.  If there are no atoms to
the right of this vacancy, STOP. \label{algo1step1}

\item Let $G$ be the set of atoms that are to the right of $v$.
Shift all the atoms in $G$ left by one step. GOTO \ref{algo1step1}.

\end{enumerate}

We say that a vacancy is at the right side if all the atoms are to
its left.  Since there are at most $n$ vacancies and each
operation can take one vacancy to the right side, the number of
operations needed is $n$. On average there are $(1-p_{occ}) n$ vacancies,
so the expected number of shifts required is $(1-p_{occ}) n$. Also,
if there are exactly $v$ vacancies that are not
on the right side, then the number of operations needed will be
$v$.

Clearly, the cost of this algorithm in the average case can be lowered by
finding the center of mass of the set of atoms and compacting them
around this.  However, this procedure does not improve the worst case cost.

\subsection{Two Dimensional Lattice}

\begin{figure}
\includegraphics[width=3.0in] {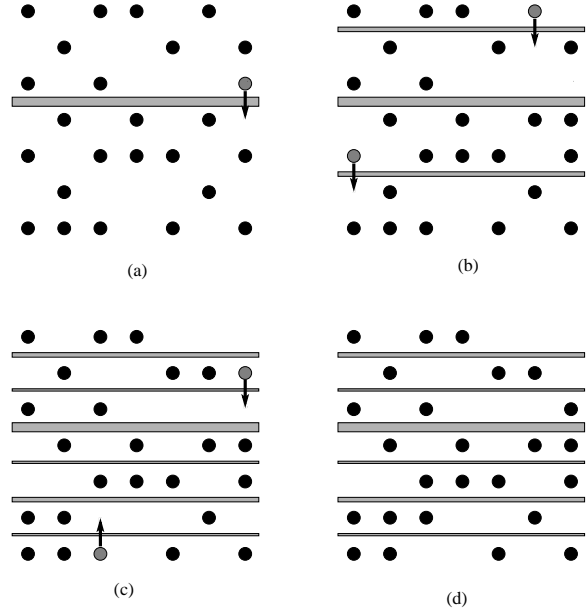}
\caption{Here we show the movement of atoms during the balancing
process for a 7x7 lattice.  Figure (a) shows the first halving and
balancing, corresponding to the 0th level of the binary tree from
Fig. \ref{fig:2dtree}. Figure (b) shows the balancing
corresponding to the second level of the tree.  Figure (c) shows
the balancing corresponding to the third level of the tree. Figure
(d) is the final result where all the rows are balanced. The
horizontal shaded rectangles show the partitioning of the lattice
into halves. The grey circles represent atoms that are to be moved
in the corresponding step.  The black circles represent atoms that
do not move in that step.}
\label{fig:bal}
\end{figure}

\begin{figure}
\includegraphics[width=3in] {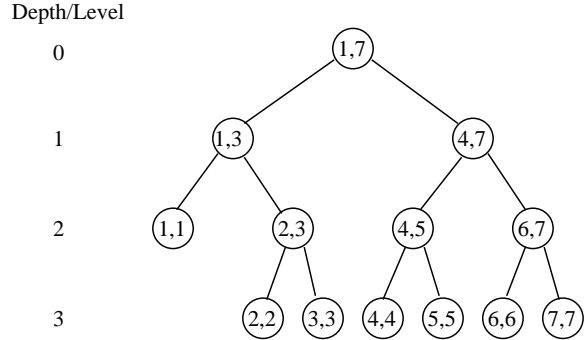}
\caption{The figure shows the binary tree describing the recursion
of BALANCE when it is given $S_{1,7}$ i.e. a 7x7 lattice viz. the
one considered in Fig. \ref{fig:bal}. The nodes of the tree show
the starting and ending rows of the sublattices balanced in each
step.  The root of the tree has depth or level 0. The maximum
depth is $\lceil \log_2 n\rceil $ which is $3$ in this case, since
$n=7$. Note that at the last level, i.e. at the leaves of the
tree, all the lattices consist of one row only. Hence no balancing
needs to be done at this point.  Thus the total number of balancing steps is $\lceil
\log_2 n\rceil$.}
\label{fig:2dtree}
\end{figure}

\begin{figure}
\includegraphics[width=3in] {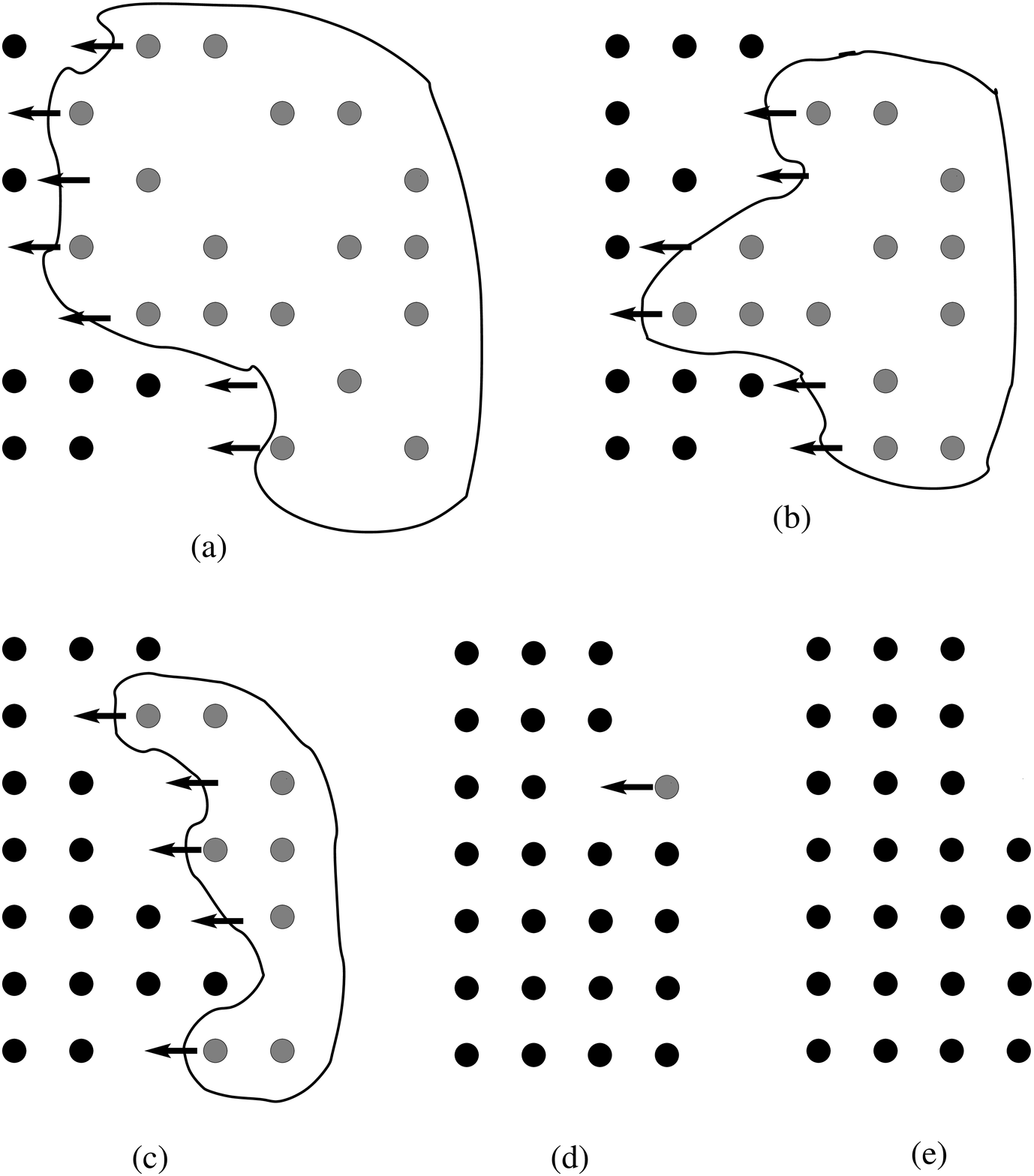}
\caption{Here we show the movement of atoms during the
row-compacting process for the  7x7 balanced lattice obtained at
the end of the balancing shown in Fig. \ref{fig:bal}. 
The grey circles represent atoms that
are to be moved in the corresponding step. The black circles
represent atoms that do not move in that step.}
\label{fig:compact}
\end{figure}

We consider a finite two dimensional square lattice $L_{n,n}$ with
$N=n^2$ sites. We denote a sublattice as $S_{i,j}$  where $(1 \le
i\le j \le n)$. $S_{i,j}$ consists of all rows between and including
rows $i$ and $j$.  Let $N(S_{ij})$ denote the total number of
atoms in the sublattice $S_{i,j}$.  Let $n_k$ represent the number
of atoms in the row $k$.

In this case the compacting procedure has two parts:

\begin{enumerate}
\item $^2$BALANCE $(S_{1,n})$: The atoms are first moved so that they are
equally distributed among all the rows.  The general idea is to
divide the lattice into two halves, e.g. the top half and the
bottom half (Fig. \ref{fig:bal}).  Then, atoms are moved from the top half to the
bottom half or vice versa, so that each half contains the same
number of atoms (a difference of one atom is allowed when the
total number of atoms is odd).  After this,  each half of the
lattice is given as input to this procedure again, thus
recursively balancing all the rows.  Fig. \ref{fig:bal}
illustrates this procedure for a 7x7 lattice. A more detailed
description of the algorithm can be found in appendix
\ref{app:2dbal}. The recursive procedure can be summarized 
graphically by a binary tree structure (Fig. \ref{fig:2dtree}),
as described below, in order to find its cost.

\item $^2$ROW-COMPACT: The atoms in each row are then compacted
as in the 1-D case, except here all rows can be compacted in
parallel since mobile atoms can be moved together as a
group. Fig. \ref{fig:compact} illustrates this procedure for a 7x7
lattice that has already been balanced according to Fig. \ref{fig:bal}.

\end{enumerate}

Now we analyze the cost of the entire procedure.  First we analyze
the cost of ROW-COMPACT.  It is the same as doing the 1-D compacting on each
row, but we can do all the rows in parallel so the worst case cost
is $n$ \cite{Note:2} (there being only $n$ sites per row, thus
requiring at most $n-1$ moves to get an atom to the left).  The
average case cost is again $(1-p_{occ}) ~n$, since there are on average
$(1-p_{occ}) n$ vacancies per row.  Additionally, if $v$ is the maximum
number of internal vacancies amongst all rows, then the cost
is $ v $ since we have to move all the vacancies beyond the last
atom on the right.

Let us now consider the cost of the balancing procedure.  We can represent the
recursion by a binary tree of depth $\lceil\log_2 n\rceil$ as
shown in Fig. \ref{fig:2dtree}.  The nodes of the tree are
labelled by the first and last rows of the sublattice $S_{i,j}$
passed as input to BALANCE. The set of nodes at the same depth $d$
are said to be at level $d$. The root is defined to be at depth 0.
During each recursive step, the number of rows in the sublattice
is halved, so the depth of the tree is $\lceil\log_2 n \rceil$
\cite{Note:3}. The total number of shifts required is the sum of the
number of shifts required at each level of the tree. Notice
that the number of atoms to be moved during a balancing step is
not a problem since they can move in parallel as a group. We
basically need to count the number of shifts that the atoms have to
make at each level.  For a sublattice of $k$ rows the
atoms only need to make $\lceil k/2 \rceil$ shifts at most,
since that is the largest number of rows of the two halves that
are being balanced. At depth $d$ the number of rows in a
sublattice is at most $\frac{n}{2^{d+1}}$ \cite{Note:4}.  There are $2^{d+1}$ such
sublattices to be balanced at depth $d$.  Now, the balancing of
all these sublattices can be done in parallel: first move all the
atoms in all these sublattices that need to be moved up.  This
takes at most $\frac{n}{2^{d+1}}$ shifts.  Then shift all the atoms
in all the sublattices that need to be moved down.  This also
takes at most $\frac{n}{2^{d+1}}$ shifts. Thus the total number of shifts
required is $\frac{n}{2}+
 (\frac{n}{2} +
 \frac{n}{4} ... (\lceil\log_2 n\rceil \mathrm{terms}))\: \le \frac{n}{2} +
 \frac{n}{2}
\sum_{n=0}^{\infty} (\frac{1}{2})^n = \frac{3}{2}n$, where the
first term in the summation is only $n/2$ (not doubled) because
there is only one direction to move the atoms the first time the
lattice is halved. Also note that one would expect to be able to balance 
each level of the binary tree in one
step on average. This would imply that the average complexity of
balancing is $\approx \log_2 n$ shift operations.  Our preliminary
simulations suggest that this is indeed the case.

Putting together the above component of the cost analysis we see that in the two
dimensional lattice with $N=n^2$ sites, the compacting procedure
takes  at most $n + \frac{3}{2} n = \frac{5}{2} n = \frac{5}{2}
\sqrt{N}$ steps (ignoring terms $O(\log N)$).

\subsection{Three Dimensional Lattice}

\begin{figure}
\includegraphics[width=6in] {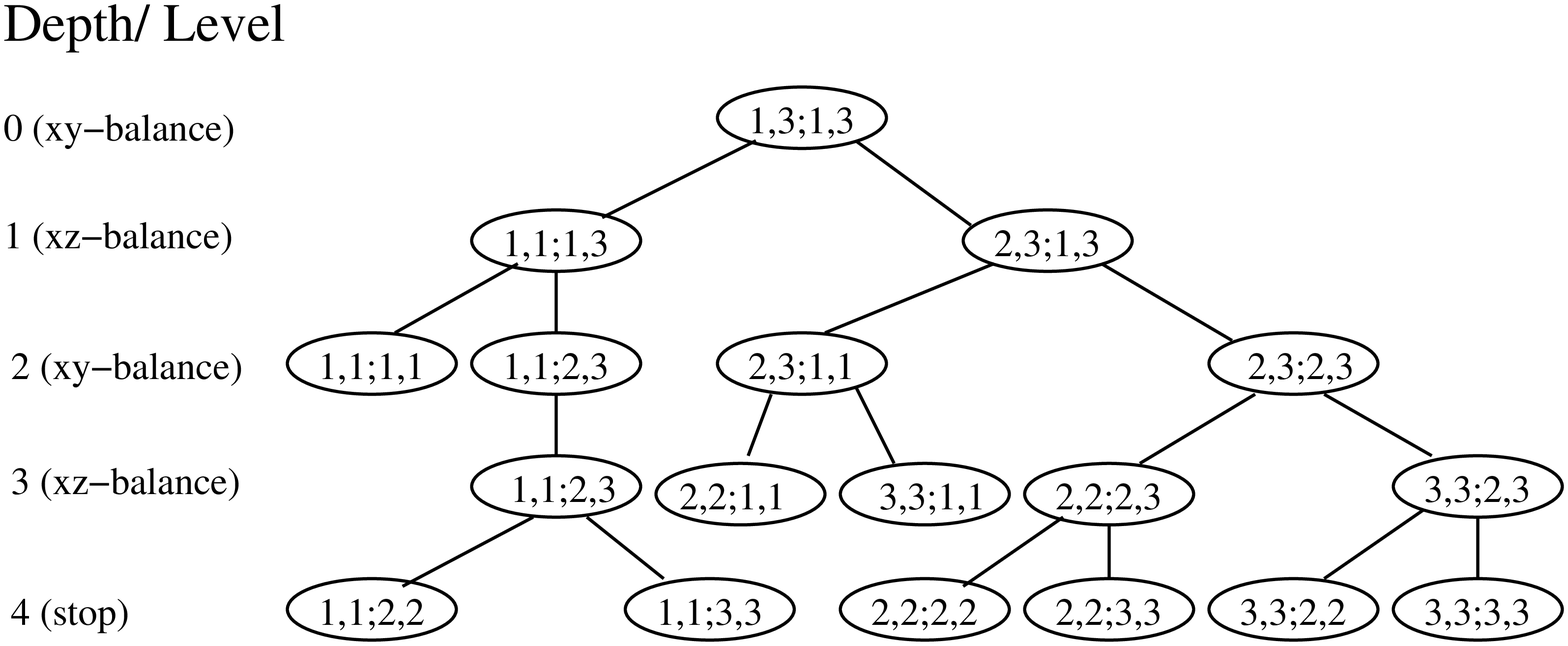}
\caption{The binary tree describing the recursion
of BALANCE when it is given $S_{1,3;1,3}$ as input, i.e. $n=3$.
The nodes of the tree show the indices of the sublattices balanced
in each step.  The root of the tree has depth or level 0. The
maximum depth is $2 \lceil \log_2 n\rceil $ which is $4$ in this
case since $n=3$ to the even levels correspond to balancing w.r.t.
to the first set of indices (halving plane is parallel to $xy$
plane), while the odd levels correspond to balancing w.r.t. the
second set of indices (halving plane is parallel to the $xz$
plane).  Note that at the last level, i.e. at the leaves of the
tree, all the lattices consist of one row only hence no balancing
needs to be done.  Thus the number of non-trivial balancing steps
is $2 \lceil \log_2 n\rceil$.}
\label{fig:3dtree}
\end{figure}

We consider a three
dimensional cubic lattice $L_{n,n,n}$ with $N=n^3$ sites. The
compacting procedure for this three dimensional case 
is a generalization of that for the two
dimensional case.  One key difference is that the notion of rows
needs to be defined.  Say, the lattice is in the positive octant
of the coordinate system, and its three sides coincide with the
axes. Taking the lattice spacing to be the unit along the axes,
the lattice sites are given by the coordinates $(i,j,k)$ with
$1\le i,j,k \le n$, where the first coordinate is the $x$
coordinate, the second coordinate is the $y$ coordinate and the
third coordinate is the $z$ coordinate.  We define row $R_{i,j}$
to be the ordered set of lattice sites $\{(l,i,j) \: | \:1\le l
\le n\}$. The $l^{\mathrm th}$ position in the row $R_{i,j}$
corresponds to the site $(l,i,j)$.

Define a sublattice $S_{i,j;k,l}$ with $(1\le i\le j \le n \: ,
1\le k\le l \le n$) to consist of the set of lattice sites
$\{(a,b,c) \; | \: 1\le a \le n, \: i\le b \le j ; k\le c\le l\}$.
Let $N(S_{i,j;k,l})$ denote the total number of atoms in the
sublattice $S_{i,j;k,l}$.

The lattice compacting procedure consists of
\begin{enumerate}
\item $^3$BALANCE($S_{1,n;1,n}$,0): The atoms are moved so that they
are equally distributed amongst the $n^2$ rows.  The second
parameter in the function call corresponds to the current
recursion depth.  The general idea is the same as in the case of
the two dimensional lattice. The difference is now that we
alternately balance the two halves made by planes parallel to the
$xy$ plane and the $xz$ plane, using the recursion depth $d$ to
implement this alternation of balancing.

\item $^3$ROW-COMPACT: The atoms in each row are then compacted in a
fashion similar to the 1-D case, except now the operations for all
the rows can be done in parallel since we can move a three
dimensional group of atoms together in a single shift.

\end{enumerate}

Now we analyze the cost of the entire procedure.  First we analyze
ROW-COMPACT.  The analysis is identical to that for the two
dimensional case.  For the sake of completeness we mention the
results here. The worst case cost is $n$. The average case cost is
$(1-p_{occ}) n$. Additionally, if $v$ is the maximum number of internal
vacancies amongst all rows then the cost is $ v $.

The analysis of balancing for the three dimensional lattice is similar to
the two dimensional case. During each recursion step the lattice
is halved along the $xy$ or $xz$ plane so the depth of the
recursion is $2 \lceil\log_2 n\rceil$. We can represent the
recursion by a binary tree of depth $2 \lceil\log_2 n\rceil$,
where even levels correspond to halving the sublattice parallel to
the $xy$ plane and odd levels correspond to halving the
sublattice parallel to the $xz$ plane. The sum of the number of
shifts required for each level of the tree gives us the total
number of shifts required during the execution of the
algorithm. We will do the counting separately for the odd and even
levels.  Since mobile atoms can be moved in parallel, we only
need to count the number of shifts that the atoms have to make
during each balancing step.  For a sublattice of $k$ rows
along the direction in which we halve the sublattice atoms, we
only need to make $\lceil k/2 \rceil$ shifts. At depth $2d$
and $2d+1$ the relevant number of rows in the sublattice is
$\frac{n}{2^{d+1}}$ (taking the root to be at depth $0$) \cite{Note:5}.
As in the 2-D case, we parallelize the required shifts for all the
sublattices at a given depth $k$, by doing all the shifts in
the same direction in parallel. Since there are two directions, the
number of shifts required at each depth is twice that required for
a single sublattice at that depth. Then the total number of shifts at the even
levels is $n/2 + n/2 + n/4 + n/8 ... (\lceil\log_2 n\rceil \mathrm{terms}))\: 
\le n/2 + n/2 \sum_{n=0}^{\infty} (1/2)^n = 3/2n$. The first
term in the sum is only $n/2$ (not doubled) because there is only
one direction to move the atoms the first time the lattice is
halved. For the odd levels a similar argument holds, except that
the first term is $n$ because there are two sublattices at that
stage and balancing them may require moving atoms in opposite
directions. Then the total number of moves required at the odd
levels is $n + n/2 + n/4 + n/8  ... (\lceil\log_2 n\rceil \mathrm{terms}))\:
\le n \sum_{n=0}^{\infty} (1/2)^n = 2n$.  Adding together the total
moves at the even and odd levels gives us the total number of
moves required: $3/2 n + 2 n = 7/2 n $.  
Thus, in the three dimensional lattice with $N=n^3$ sites, the
lattice can be compacted in at most $7/2 n+ n = 9/2 n = 9/2
\sqrt[3]{N}$ steps (ignoring terms $O(\log N)$).
As in the case of the two
dimensional lattice, one would expect that the balancing at each
level of the binary tree can be typically done in one step. This
would imply that the average cost of balancing is $2
\lceil\log_2\rceil n$ steps. 
However, we do not have a rigorous proof of this estimate.

\section{Conclusion}

In conclusion, we have presented an experimentally viable scheme to perfectly initialize a quantum computer based on neutral atoms trapped in an optical lattice of large lattice constant. The proposed compacting scheme allows preparation of a perfect optical lattice of orthorhombic structure, where each lattice site is occupied with a single atom in a specific internal level and motional ground state. We have proposed physical mechanisms for realization of the two elementary compacting operations, the flip of the internal state of atoms and their shift within the lattice structure, and have analyzed their efficiency in some detail. Particular attention was devoted to the study of the motional heating during the flip and shift operation.
Mechanisms to control this heating were proposed that are based on the analytical and numerical solutions of the atomic dynamics on the time-dependent lattice potential. Our analysis of the complexity of the compacting process demonstrates its scalability with the size of the quantum computer.

We point out that all of the components discussed here are feasible with current technology. 
Our results show that we can achieve a perfect optical lattice in less than one second starting from a half filled lattice of approximately 8000 sites. 
The procedure can also be used to complement other proposals that provide near
complete but not perfect filling, provided they are carried out in a site addressable optical lattice.
Furthermore, the lattice compacting techniques proposed here can be adapted to the execution of quantum logic gates in site-addressable optical lattices.
The control of individual atoms in an optical lattice envisioned here would set the stage for the construction of a neutral atom quantum computer.

\begin{acknowledgments}
We thank Ronald de Wolf and Kenneth Brown for helpful discussions.
The effort of the authors is sponsored by the Defense Advanced Research Projects Agency
(DARPA) and the Air Force Laboratory, Air Force Material Command, USAF, under agreement
numbers F30602-01-2-0524. D.S.W. also acknowledges support from the National Science Foundation.

\end{acknowledgments}

%%%%%%%%%%%%%%%%%%%%%%%%%%%%%%%%%%%%%%%%%%%%%%%%%%%%%%%%%%%%%%%%%%%%%%%%%
%%%%%%%%%%%%%%%% APPENDICES START HERE %%%%%%%%%%%%%%%%%%%%%%%%%%%%%%%%%%
%%%%%%%%%%%%%%%%%%%%%%%%%%%%%%%%%%%%%%%%%%%%%%%%%%%%%%%%%%%%%%%%%%%%%%%%%
\appendix

\section{Method of numerical simulation}
\label{App:Methods}

In this appendix, we summarize the two parts of our approach to
numerical simulation of atomic motion on an optical potential.
First, we introduce the Fourier grid method, a discretization of
the Schr$\ddot{\text{o}}$dinger equation suitable for computational
implementation. Second, an accurate and efficient
approach, based on a Chebychev polynomial expansion of the time
evolution operator, is presented. In both paragraphs, we discuss the
advantages and disadvantages of the methods.

\subsection{Fast Fourier grid method}
\label{sec:grid}
A combination of continuous and discrete degrees of freedom characterize
full scale dynamics of most quantum mechanical system. The
discrete degrees, at least in cases of smaller dimensionality, are often
easily handled, but the continous degrees of freedom require that some form of
discretization scheme be used.

Finite differences introduced grid methods into quantum molecular dynamics
simulations, but, since they suffered from poor accuracy and slow
convergence, they were soon replaced by other
methods. The Fourier method does not
suffer from the same shortcomings, and allows for rapid convergence
and efficient sampling of system phase space.~\cite{kosloff96,kosloff98}

The Fourier basis functions, $\theta_k = e^{2\pi ikx/L}$, form an
orthogonal basis, and can be used to represent a system wavefunction
or operator to arbitrary accuracy.
\begin{equation}
\ket{\Psi(x)} = \sum_{k=-N}^{N-1} a_k \, \theta_k
\end{equation}

\begin{eqnarray}
\ket{\Psi(x)}' &=& O \ket{\Psi(x)} \nonumber \\
\ket{\Psi(x_k)}' &=& \sum_{l=1}^{2N} O_{kl} \ket{\Psi (x)}
\end{eqnarray}
For any non-local system operator, {\em e.g.} the kinetic energy
operator in the position representation, the evaluation of the
operator, $O$, scales as the square of the number of grid
points. In contrast, the potential energy operator, $U(x)$,
which is local in position space, scales linearly with the number of
grid points and is thus simple to
evaluate. The advantage of the Fourier method is
that the kinetic energy operator, $T(p)$, can also be
evaluated as a diagonal operator if the wavefunction is expressed in
the momentum basis. Consequently, the action of a given Hamiltonian,
$H(x,p)=T(p)+U(x)$, can be evaluated as
\begin{equation}
H \Psi(x) = \frac{1}{2N} \sum_{m=-N}^{N-1}
e^{\frac{2\pi impx}{L}} T(p) \sum_{n=-N}^{N-1} e^{-\frac{2\pi
inpx}{L}} \Psi(x) + U(x) \Psi(x)
\end{equation}
The first term describes the action of the kinetic energy
operator. It explicitly takes into account the two transformation steps:
i) the Fourier transform into the momentum representation followed by
application of the diagonal kinetic energy operator; and ii)
the inverse transform back into the position representation. Additonally,
to facilitate a direct application to physical problems,
the Fourier transform is defined on the finite interval $[-L/2, L/2]$,
giving rise to the $2\pi/L$ term in the exponent.

The Fourier method is limited to bandwidth limited functions 
that decay exponentially fast as the argument approaches
infinity. For such functions the Whittaker-Kotel'nikov-Shannon
theorem ensures that, with a finite number of sampling points, the
value of the function can be determined to any desired
accuracy.~\cite{kosloff93} In the case of time-dependent
problems, the choice of sampling points has to ensure
that the wavefunction remains confined at all time to the space defined by 
these points.

\subsection{Chebychev propagation method}
\label{sec:cheb}
The Chebychev propagation algorithm employs the orthogonal Chebychev
polynomials to approximate the quantum mechanical time evolution
operator, $U(t)$, or any other operator functional. In contrast to
other interpolation methods, the
Chebychev method does not suffer from inherent instabilities that limit
the order of the expansion, and consequently also the attainable
accuracy.~\cite{kosloff94} Moreover, the interpolation error is always
uniform, and can, due to the possibility of very high order
expansions, be reduced to the order of the computer's numerical
precision.

Because Chebychev polynomials are defined on the interval
$[-1,1]$, it is necessary to map the Hamiltonian onto this
interval using the linear transformation
\begin{equation}
H'=2\frac{ H-E_{min}I }{E_{max}-E_{min}} - I.
\end{equation}
Here $H$ is the Hamiltonian of the system,
$E_{max}$ and $E_{min}$ are the estimates of the largest and smallest
energy eigenvalues of $H$, and $I$ is the identity operator.
For the next step of the Chebychev propagation algorithm, the Chebychev
expansion coeffecients are calculated:
\begin{eqnarray}
b_0=J_0 \left( \frac{\Delta E \cdot \Delta t}{2 \hbar} \right)
\theta (\Delta t) \\
b_k=2i^k J_k \left( \frac{\Delta E \cdot \Delta t}{2 \hbar} \right)
\theta (\Delta t).
\end{eqnarray}
Here $b_k$ is the expansion coefficient for the Chebychev polynomial of
order k, $T_k(H')$. $\theta(\Delta t)$ is a global phase
factor, $\theta(\Delta t)=exp(-(i/\hbar)(\Delta E/2+E_{min})\Delta
t)$, arising from the mapping step. The appearance of the
Bessel function, $J_k$, is attractive since it provides an automatic
cut off point for the expansion. As the order of the expansion
increases beyond the size of the argument, $\frac{\Delta E \cdot \Delta
t}{2 \hbar}$, the associated expansion coeffecients will approach
zero exponentially fast.~\cite{kosloff94} The time evolution operator
can now be approximated as
\begin{equation}
U(\Delta t)' = \sum_{k=0}^{N} b_k T_k \left( H' \right)
\end{equation}
Here $b_k$ is the expansion coefficient for the k'th order Chebychev
polynomial, $T_k(H')$. More efficient than
calculating and applying the full expansion to the system
wavefunction, however, is recursively updating the system
state. This recursive approach is particularly attractive since it
permits the expansion to be terminated when the desired precision has
been attained. Recursion is carried out as follows:
\begin{eqnarray}
\ket{\Psi(t+\Delta t)}_0 &=& \ket{\Psi(t)} \nonumber \\
\ket{\Psi(t+\Delta t)}_1 &=& H' \ket{\Psi(t)} \nonumber \\
\ket{\Psi(t+\Delta t)}_k &=& 2 H' \ket{\Psi(t+\Delta
t)}_{k-1} - \ket{\Psi(t+\Delta t)}_{k-2}.
\end{eqnarray}

\section{ 2-D BALANCE}
\label{app:2dbal}

$^2$BALANCE$(S_{i,m})$ see Sec. \ref{Sec:Complexity} for definitions.

\noindent{\bf INPUT:} $S_{i,j}$ \\

\noindent {\bf GOAL:} Balance the rows in $S_{i,j}$ so that
$\forall \; k,l$  we have  $|n_k - n_l| \le 1$, where $i\le k,l
\le j$. Equivalently, each row has at least $n_{min} = \lfloor
N(S_{i,j})/(j-i+1)\rfloor $ atoms. \\

\noindent {\bf ALGORITHM}
\begin{enumerate}

\item Let $l = (j-i + 1)$ denote the number of rows in $S_{i,j}$.
If $l = 1$  RETURN.

\item Let $m = i + \lfloor l/2 \rfloor$ be a middle row of
$S_{i,j}$.  For balancing, we need to have $n_{req} = (m-i+1)
n_{min}$ atoms in the sublattice $S_{i,m}$.

\item If $(N(S_{i,m}) > n_{req})$, shift $(N(S_{i,m}) - n_{req})$
atoms down from $S_{i,m}$ to $S_{m+1,j}$

\item If $(N(S_{i,m}) < n_{req})$, shift $(n_{req} - N(S_{i,m}))$
atoms up from $S_{m+1,j}$ to $S_{i,m}$.

\item $^2$BALANCE$(S_{m+1,j})$.

\end{enumerate}

\section{3D-BALANCE}
\label{app:3dbal}

$^3$BALANCE$(S_{i,m;k,l},d+1)$ see Sec. \ref{Sec:Complexity} for definitions.

 \noindent {\bf INPUT:} Sublattice  $S_{i,j;k,l}$
and recursion depth $d$.

\noindent{GOAL:} Balance the rows in $S_{i,j;k,l}$ so that
$\forall_{a,b}$ such that $i\le a,b \le j$ and $\forall_{c,d}$
such that $k \le c,d \le l$  we have $|n_{a,c} - n_{b,d}| \le 1$,
where $n_{a,c}$ represents the number of atoms in the row
$R_{a,c}$. Equivalently, each row should have at least $n_{min} =
\lfloor \frac{N(S_{i,j;k,l})}{(j-i+1)(l-k+1)}\rfloor $ atoms.\\

\noindent {\bf ALGORITHM:}
\begin{enumerate}

\item If $i = j$ and $k = l$ then RETURN.\\
Only one row so no balancing needed.

\item If $d$ is odd, goto \ref{xzbalance} else if $i = j$ $^3$BALANCE$(S_{i,j;k,l},d+1)$. RETURN.

\item If sublattice $S_{i,j;k,l}$ contains only one $xy$ plane
(i.e. $i = j$), then goto \ref{xzbalance}.

\item \label{xystart} Let $l = j-i + 1$ be the number of planes
parallel to the $xy$ plane. Then $m = i + \lfloor l/2 \rfloor$
defines a middle plane parallel to $xy$ plane.

\item Let $n_{req} = n_{min} (m-i+1) (l-k+1)$ be the minimum
number of atoms that are required to be in $S_{i,m;k,l}$.

\item If $(N(S_{i,m;k,l})> n_{req})$, shift $(N(S_{i,m;k,l}) -
n_{req}) $ atoms from $S_{i,m;k,l}$ to $S_{m+1,j;k,l}$.

\item If $(N(S_{i,m;k,l}) < n_{req})$, shift $n_{req} -
N(S_{i,m;k,l})$ atoms from $S_{m+1,j;k,l}$ to $S_{i,m;k,l}$.

\item $^3$BALANCE$(S_{m+1,j;k,l},d+1)$ \label{xyend}.

\item RETURN

\item \label{xzbalance} If $k = l$ $^3$BALANCE$(S_{i,j;k,l},d+1)$. RETURN.

\item Do steps \ref{xystart} to \ref{xyend}, replacing indices
$(i,j)$ with indices $(k,l)$.

\item RETURN

\end{enumerate}

\newpage
\bibliographystyle{apsrev}

\begin{thebibliography}{37}
\bibitem[{Not({\natexlab{a}})}]{Note:1}
\bibinfo{note}{We employ standard computer science notation that $O(n)$ means
  scaling at most linear with $n$ for large $n$.}
\bibitem[{Not({\natexlab{b}})}]{Note:2}
\bibinfo{note}{The precise value is $n-1$ but we ignore the constant for
  clarity and since we are more concerned with the scaling behaviour.}
\bibitem[{Not({\natexlab{d}})}]{Note:4}
\bibinfo{note}{Actually this should be $\lceil \frac{n}{2^{d+1}} \rceil$, but
  for clarity and simplicity we ignore this. It only contributes an extra term
  of $\lceil\log n\rceil$ to the worst case cost, which is negligible compared
  to the other terms.}
\bibitem[{Not({\natexlab{e}})}]{Note:5}
\bibinfo{note}{Actually this should be $\lceil \frac{n}{2^{d+1}} \rceil$, but
  for clarity and simplicity we ignore this. It only contributes an extra term
  of $2 \lceil\log n\rceil$ to the worst case cost, which is negligible
  compared to the other terms.}
\end{thebibliography}

\end{document}